\documentstyle[preprint,prl,aps]{revtex}
\begin{document}

\tightenlines
\preprint{SNUTP 97-136, quant-ph/9709046}
\draft
\title{
Interference phenomena in the photon production between two 
oscillating walls
}
\author{
Jeong-Young Ji\footnote{Electronic address: 
jyji@phyb.snu.ac.kr},  
Hyun-Hee Jung\footnote{Electronic address: hhjung@gmc.snu.ac.kr},
Kwang-Sup Soh\footnote{Electronic address: kssoh@phya.snu.ac.kr}
}
\address{Department of Physics Education, Seoul National University, 
Seoul 151-742, Korea
}
\maketitle

\begin{abstract}
We study the photon production in a 1D cavity whose left and right walls 
oscillate with the frequency $\Omega_{L} $ and $\Omega_{R} $, 
respectively.  For $\Omega_{L} \neq \Omega_{R} , $ the number of generated 
photons by the parametric resonance is the sum of the photon numbers 
produced when the left and the right wall oscillates separately. But for 
$\Omega_{L} = \Omega_{R} $, the interference term proportional to 
$\cos \phi $ is found additionally, where $\phi $ is the phase difference 
between two oscillations of the walls.
\end{abstract}


\narrowtext

The standard Casimir effect~\cite{Casimir48} indicates the presence of 
vacuum fluctuations of the electromagnetic field. The modified boundary 
conditions of the field, which characterize the dynamical situation, 
change zero-point vacuum fluctuations inside a perfectly reflecting cavity 
and give photon productions~\cite{Moore70}.
This phenomenon has been extensively
studied when the one of the walls oscillates~\cite
{DoKM90,Jaekel,SaSW94,Law94,ColeS95,MeplanG96,Do95,DoK96,KlimovA97,J}. 
Photon productions by parametric resonance when the wall of the cavity 
oscillates with one of the resonant field frequencies, give one a 
possibility to observe significant effects under experimental situations 
and to obtain quantitative
results~\cite{Do95,DoK96,KlimovA97}. 
 
Recently, Lambrecht {\it et al.}~\cite{LambJR96} studied the radiation 
emitted by two oscillating walls using the scattering approach. Therein, 
boundaries oscillate symmetrically with respect to the center of the 
cavity or globally with the length of cavity kept constant. Resonance 
enhancement occurs when the oscillating walls have even integer multiples 
of the fundamental frequency of the cavity for the former and odd integer 
multiples for the latter. If we adopt the perturbation approach of 
Ref.~\cite{J} the above results can be easily obtained. Moreover, we can 
further proceed to a more general situation where the walls oscillate with 
different frequencies and with a phase difference. 

In this Rapid Communication, we shall present the interference phenomena 
in the number of parametrically generated photons when two walls oscillate 
with the same frequency but with a phase difference. Further, we shall 
show that the result of Ref.~\cite{LambJR96} can be explained by a 
constructive interference or a destructive interference. For this purpose, 
we shall calculate the number of produced photons by the parametric 
resonance in a cavity with two oscillating walls. For the different 
frequencies, there is no interference and the number of generated photons 
by the parametric resonance is the sum of the photon numbers produced when 
the left and the right wall oscillates separately.

We start with a field operator $A(x,t) $ associated with a vector 
potential which satisfies the one-dimensional wave equation $(c=1) $
\begin{equation}
\frac{{{\partial^{2}}  A}}{\partial t^{2}} - \frac{{{ \partial^{2}}  
 A}}{\partial x^{2}}=0 .
\label{WEQ}
\end{equation}
Following Refs.~\cite{DoKM90,Law95}, we expand the field operator as
\begin{equation}
A (x,t) = \sum_{n} 
[b_{n} \psi_{n} (x,t) + 
b_{n}^{\dagger}\psi_{n}^{*}(x,t) ]
\label{Aps}
\end{equation} 
where ${b_{n}}^{\dagger}   $ and $b_{n} $ are the usual creation and 
annihilation operators. Here $\psi_{n} (x,t) $ is the mode function which 
vanishes at the left and the right walls which are located at $x = L(t)  $ 
and $x=R(t) $ respectively, i.e. 
$\psi_{n} (L(t), t) = 0 = \psi_{n} (R(t),t) $. Let us introduce the 
instantaneous mode basis function
\begin{equation}
\varphi_{k} (x, t) = \sqrt{ \frac{2}{R - L}} \sin \frac{{k \pi ( x - L 
 )}}{R - L},
\end{equation}
to evolve
\begin{equation}
\psi_{n} (x,t) = \sum_{k=1}^\infty Q_{nk} (t) \varphi_{k} (x,t)
\label{PQ}
 \end{equation}
where $k $ is a positive integer. Now we consider the small motions of 
the walls $(\epsilon  \sim \epsilon_{L} \sim \epsilon_{R} \ll  1) $
\begin{equation}
L(t) = \Lambda \epsilon_{L} ( t )
\end{equation}
and
\begin{equation}
R(t) = \Lambda [ 1 + \epsilon_{R} (t) ] ,
\end{equation}
where $\Lambda  $ is the distance of two walls in the static situation 
and $\epsilon $ is small parameter which characterizes the small deviation 
of the walls from the initially static position. Considering only to the 
first order in $\epsilon $, from (\ref{WEQ}) and the orthogonality of the 
mode functions we have an infinite set of coupled differential equations 
\begin{eqnarray}
{\ddot{Q}}_{nk} 
&=& 
- {\omega_{k} }^2 Q_{nk}
+ 2 \omega_{k}^{2} 
[ \epsilon_{R} (t) - \epsilon_{L} (t) ] Q_{nk} 
\nonumber \\
&& + 
2 \sum_{j} G_{kj} {\dot{Q}}_{nj}
+ \sum_{j} \dot{G}_{kj} Q_{nj} 
+ O( \epsilon^{2} )
\label{EOM}
\end{eqnarray}
where 
\begin{equation}
G_{kj} = \frac{{g_{kj}^R \dot{R} - g_{kj}^L \dot{L} }}{\Lambda}
\label{G}
\end{equation}
and $\omega_{k} $ is the mode frequency for the static walls:
\begin{equation}
\omega_{k} = \frac{{k \pi}}{\Lambda} .
\label{wk}
\end{equation} 
Here we used the following relations
\begin{equation}
g_{jk}^L \equiv 
-(R-L) \int_{L}^{R} \varphi_{k} \frac{{ \partial \varphi_{j} }}{ \partial 
 L } dx 
= \frac{{2jk}}{k^{2} - j^{2}}
\label{gL}
\end{equation}
and
\begin{equation}
g_{jk}^R \equiv
(R-L) \int_{L}^{R} \varphi_{k} \frac{{ \partial \varphi_{j} }}{ \partial 
 R } dx 
=(-1)^{j+k} \frac{{2jk}}{k^{2} - j^{2}} .
\label{gR}
\end{equation}

Now we consider the special motion of the walls where the left and the 
right walls oscillate according to 
\begin{equation}
\epsilon_{L} (t) = \epsilon a_{L} \sin ( \Omega_{L} t + \phi_{L} )
\end{equation}
and
\begin{equation}
\epsilon_{R} (t) = \epsilon a_{R}  
\sin ( \Omega_{R} t + \phi_{R} ) ,
\end{equation}
where, without loss of generality, we set 
$\phi_{L} = \phi  ~ {\rm and} ~ \phi_{R} = 0 $. 

Introducing the new dynamical variable~\cite{J}
\begin{equation}
X_{n, k \mp} = \sqrt{ \frac{\omega_{k}}{2}} 
\left( Q_{nk} \pm i \frac{{\dot{Q}}_{nk}}{\omega_{k}} \right),
\label{X:QP}
\end{equation}
and using the vector notation
\begin{equation}
\vec{X}_{n}(t)  = ( 
X_{n,1-} ,  X_{n,1+} ,  X_{n,2-} ,  \cdots )^{T} ,
\end{equation}
Eq. (\ref{EOM}) is transformed to the following first order differential 
equation 
\begin{equation}
\frac{d}{dt} \vec{X}_{n} (t) = V^{(0)} \vec{X}_{n}(t) 
+ \epsilon V^{(1)} \vec{X}_{n}(t) \label{deXn}
\end{equation}
where $V^{(0)} $ and $V^{(1)} $ are matrices given by
\begin{equation}
V_{k \sigma, j \sigma'}^{(0)} = i \omega_{k} \sigma \delta_{kj} 
 \delta_{\sigma \sigma'}
\end{equation}
and
\begin{equation}
V_{k \sigma, j \sigma'}^{(1)} = 
\omega_{1} \sum_{s = \pm}
\left(
v_{k \sigma, j \sigma'}^{Rs} e^{s i \Omega_{R} t}
- v_{k \sigma, j \sigma'}^{Ls} e^{s i \Omega_{L} t}
\right) ,
\label{V1}
\end{equation}  
where
\begin{eqnarray}
v_{k \sigma, j \sigma'}^{A s} 
&=& \sigma a_{A} e^{i s \phi_{A}} 
\nonumber \\
&& 
\times \Bigl[
\gamma_{A} g_{kj}^A \sqrt{\frac{j}{k}} 
\Bigl( \frac{{ \sigma' }}{2} + s \frac{{\gamma_{A}}}{4j} \Bigr)
-s \frac{k}{2} \delta_{kj} 
\Bigr] ,
\label{vg}
\end{eqnarray}
with $A=L, R $, and $s, \sigma,\sigma' = +,- $. Here we used 
$\Omega_{A} = \gamma_{A} \omega_{1} $ and $\omega_{k} = k \omega_{1} . $ 

Taking the following power-series expansion in $\epsilon $
\begin{equation}
\vec{X}_{n} =\vec{X}_{n}^{(0)} + \epsilon \vec{X}_{n}^{(1)} 
+ \epsilon^{2} \vec{X}_{n}^{(2)} + \cdots ,
\label{series}
\end{equation}
we have the zeroth order and first order equation
\begin{eqnarray}
\frac{d}{dt} \vec{X}_{n}^{(0)}
&=& V^{(0)} \vec{X}_{n}^{(0)} ,
\label{0th} \\
\frac{d}{dt} \vec{X}_{n}^{(1)}&=& 
V^{(1)} \vec{X}_{n}^{(0)} +V^{(0)} \vec{X}_{n}^{(1)} ,
\label{1st}
\end{eqnarray}
which can be easily integrated to give the following solutions:
\begin{eqnarray}
X_{n, k \sigma}^{(0)} (t) &=& 
\delta_{nk} \delta_{\sigma-} e^{- i \omega_{k} t} ,
\label{solX0} \\
X_{n, k \sigma}^{(1)} (t) 
&=& e^{\sigma i \omega_{k} t} 
\int_{0}^{t} dt' e^{- \sigma i \omega_{k} t' } \sum_{j, \sigma'} 
V_{k \sigma , j \sigma'}^{(1)}  X_{n, j \sigma'}^{(0)} 
\nonumber \\
&=& \omega_{1} e^{\sigma i k \omega_{1} t} 
\int_{0}^{t} dt' 
[ v_{k \sigma, n-}^{R-} 
e^{-i(\sigma k + \gamma_{R} +n) \omega_{1} t'} 
\nonumber \\
&& + v_{k \sigma, n-}^{R+} 
e^{+i(- \sigma k + \gamma_{R} - n ) \omega_{1} t'} 
- (R \leftrightarrow L) ] .
\label{solX1}
\end{eqnarray}
Using the continuity conditions at $t=T $ we can find easily the 
Bogoliubov coefficient $\beta_{nk} $ in the solution
\begin{equation}
\psi_{n} (x,t>T) = \sum_{k} \left[\alpha_{nk} \frac{{e^{-i \omega_{k} 
 t}}}{\sqrt{2 \omega_{k}}}
+\beta_{nk} \frac{{e^{i \omega_{k} t}}}{\sqrt{2 \omega_{k}}}\right] 
 \varphi_{k}(x) .
\label{psnT}
\end{equation}
For the $\epsilon $-order approximation, it is just the coefficient of 
the negative frequency $e^{+i \omega_{k} t} / \sqrt{2 \omega_{k} } $ in 
the solution 
\begin{equation}
Q_{nk} (t) = \frac{1}{\sqrt{2 \omega_{k}}} 
[X_{n,k-} + X_{n,k+} ] .
\end{equation}
By noting that $\omega_{1} T \gg  1 $ we take only the dominant terms 
which are proportional to the time 
\begin{equation}
\beta_{nk}  =  \epsilon \omega_{1} T 
( v_{k+,n-}^{R+} \delta_{k, \gamma_{R} - n} 
- v_{k+,n-}^{L+} \delta_{k, \gamma_{L} - n} ) .
\end{equation}
Using (\ref{gL}), (\ref{gR}) and (\ref{vg}), we write explicitly
\begin{equation}
\beta_{nk}  = 
\frac{1}{2} \epsilon \omega_{1} T \sqrt{k n}
\left[
(-1)^{\gamma_{R}} a_{R} 
\delta_{n, \gamma_{R} - k }
- e^{i \phi} a_{L} \delta_{n, \gamma_{L} - k}
\right] .
\label{bnk}
\end{equation}
and using $N_{nk} = | \beta_{nk} |^{2} $ we can obtain the number of 
created photons with the frequency $\omega_{k} $ which comes from the 
initial mode function with frequency $\omega_{n} $:
\begin{equation}
N_{nk}  =
N_{nk}^L + N_{nk}^R - 
2 (-1)^{\gamma_{R}} \sqrt{ N_{nk}^L } 
\sqrt{ N_{nk}^R } \cos \phi ,
\label{Nnk:LR}
\end{equation}
where
\begin{equation}
N_{nk}^A =
\left( \frac{1}{2} \epsilon \omega_{1} T \right)^{2}
k n a_{A}^{2} \delta_{n, \gamma_{A} - k }
\end{equation}
with $A=L,R. $ By summing over $n $ we finally have the total number of 
generated photons with the frequency $\omega_{k} $ 
\begin{equation}
N_{k} =
N_{k}^{L} + N_{k}^{R}  
-(-1)^{\gamma_{R}} 2 \sqrt{ N_{k}^{L} } \sqrt{ N_{k}^{R} }\cos \phi
\delta_{\gamma_{L} \gamma_{R}} ,
\label{Nk:LR}
\end{equation}
where
\begin{equation}
N_{k}^{A} =
\left( \frac{1}{2} \epsilon \omega_{1} T \right)^{2}
k ( \gamma_{A} - k ) a_{A}^{2}  ,
\end{equation}
or explicitly we have
\begin{eqnarray}
N_{k} &=& \left( \frac{1}{2} \epsilon \omega_{1} T \right)^{2}
[ k ( \gamma_{R} - k ) a_{R}^{2} 
+ k ( \gamma_{L} - k ) a_{L}^{2}
\nonumber \\
&& 
- (-1)^{\gamma_{R}} 2 k ( \gamma_{L} - k ) a_{R} a_{L} \cos \phi 
\delta_{\gamma_{L,} \gamma_{R} } ] .
\label{Nk}
\end{eqnarray}
Note that this result agrees with our previous result~\cite{J} when the 
only right-side wall oscillates $(a_{L} = 0 , ~ a_{R} = 1) $. 

Now we consider some special case where the two frequencies of walls are 
same $\Omega_{L} = \Omega_{R} = \Omega
~ ( \gamma_{L} = \gamma_{R} = \gamma ) $. If we rewrite (\ref{Nk:LR}) 
directly from (\ref{bnk}), we have the following form
\begin{equation}
N_{nk}  = 
\left( \frac{1}{2} \epsilon \omega_{1} T \right)^{2} k n
\left|
(-1)^{k+n} a_{R} - e^{i \phi} a_{L} 
\right|^{2} \delta_{n, \gamma-k} .
\label{Nnk:freq}
\end{equation}
One easily find that for $\phi = 0 \mbox{ or }  \pi $ this result 
corresponds to the situation studied in Ref.~\cite{LambJR96} [see Eq.~(12) 
in the perfect-mirror limit $(\rho \rightarrow 0) $] except for the 
time-dependence. Therein they used scattering approach and assumed the 
linear  dependence of time. For the $k$th mode photon numbers, it follows 
from (\ref{Nk:LR}) that
\begin{equation}
N_{k} =
N_{k}^{L} + N_{k}^{R} - 
(-1)^{\gamma} 2 \sqrt{ N_{k}^{L} } \sqrt{ N_{k}^{R} }\cos \phi .
\label{Nk:LR:freq}
\end{equation}
It is worth noting that this formula resembles the intensity formula of 
the double-slit interference experiment except for the factor  
$(-1)^{\gamma+1} $. The last term in (\ref{Nk:LR:freq}) is the 
interference term and the photon numbers vary as the phase difference $\phi $ 
changes.

In order to examine the interference pattern we further restrict the 
situation so that the walls oscillate with the same amplitude  
$(a_{L} = a_{R} =1) $. In this case
\begin{equation}
N_{k}  = 
\left( \frac{1}{2} \epsilon \omega_{1} T \right)^{2} 2 k ( \gamma - k )
[ 1 - (-1)^{\gamma} \cos \phi ] .
\label{Nnk:amp}
\end{equation}
When the walls oscillate with even mode frequency 
$\Omega = 2 \omega_{1} , 4 \omega_{1} , ... $, the interference effect is 
characterized by the function $1 - \cos \phi  $. Then the number of 
photons is maximal (constructive interference) at $\phi = \pi $ (the two 
walls oscillate symmetrically with respect to the center of the cavity) 
and minimal (destructive interference) at $\phi=0 $ (the walls oscillate 
together while keeping their distance constant). For 
$\Omega = 3 \omega_{1} , 5 \omega_{1} , ... $ the number of created 
photons is proportional to $1 + \cos \phi $ and maximal at $\phi=0 $ and 
minimal at $\phi = \pi $. As pointed out in Refs.~\cite{J,LambJR96} the 
photon distribution shows a parabolic spectrum and hence the maximum value 
of photon number appears at the nearest mode frequency 
$\omega_{k} = \Omega / 2 $. 

For the general cases with different frequencies of the wall 
$( \Omega_{L} \neq \Omega_{R} ) $, the interference term in (\ref{Nk:LR}) 
vanishes and we have the result of no interference
\begin{equation}
N_{k} = N_{k}^L + N_{k}^R ,
\end{equation}
that is, the number of generated photons by the parametric resonance is 
the sum of the photon numbers produced when the left and the right wall 
oscillates separately. Then there are two peaks at the nearest modes of 
frequencies $\omega_{k} = \Omega_{L} / 2 $ and $\omega_{k} = \Omega_{R} / 2 $
. 

In summary, we have calculated the number of produced photons by the 
parametric resonance in a cavity whose left and right walls oscillate with 
respective frequencies, phases, and amplitudes. For the oscillations of 
the same frequency, we have presented the interference phenomena where the 
photon number is a function of the phase difference. But for the different 
frequencies, there is no interference and the photon number is the sum of 
the photon numbers produced when the left and the right wall oscillates 
separately.

This work was supported by the Center for Theoretical Physics (S.N.U.), 
Korea Research Center for Theoretical Physics and Chemistry, and the Basic 
Science Research Institute Program, Ministry of Education Project No. 
BSRI-97-2418. One of us (J.Y.J.) was supported by Ministry of Education 
for the post-doctorial fellowship.

\end{document}